\documentclass[sigconf]{acmart}
\AtBeginDocument{%
  }

\copyrightyear{2026}
\acmYear{2026}
\setcopyright{cc}
\setcctype{by}
\acmConference[MSR '26]{23rd International Conference on Mining Software Repositories}{April 13--14, 2026}{Rio de Janeiro, Brazil}
\acmBooktitle{23rd International Conference on Mining Software Repositories (MSR '26), April 13--14, 2026, Rio de Janeiro, Brazil}
\acmPrice{}
\acmDOI{10.1145/3793302.3793347}
\acmISBN{979-8-4007-2474-9/2026/04}

\usepackage{graphicx} 
\usepackage{amsmath}
\usepackage{xspace}
\usepackage{array}
\usepackage{listings}
\usepackage{multirow}
\usepackage{url}
\usepackage{booktabs}
\usepackage{csquotes}
\usepackage[dvipsnames,table]{xcolor}
\usepackage[lined,algonl,ruled,boxed,vlined,linesnumbered]{algorithm2e}
\newcolumntype{L}[1]{>{\raggedright\let\newline\\\arraybackslash\hspace{0pt}}m{#1}}
\newcolumntype{C}[1]{>{\centering\let\newline\\\arraybackslash\hspace{0pt}}m{#1}}
\newcolumntype{R}[1]{>{\raggedleft\let\newline\\\arraybackslash\hspace{0pt}}m{#1}}
\newcommand{\toolb}{\textsc{BuCoR}\xspace}
\newcommand{\toolba}{\textsc{BuCoR-E}\xspace}
\newcommand{\toolbb}{\textsc{BuCoR-R}\xspace}

\newcommand{\codefont}[1]{\small{\texttt{#1}}\normalsize}
\newcommand{\totalcon}{88\xspace}
\newcommand{\totaltype}{21\xspace}
\newcommand{\totalResolved}{34\xspace}

\begin{document}

\lstset{language=Java,
breaklines=true,
basicstyle=\footnotesize,
escapeinside={(*@}{@*)},
belowskip=0pt,
aboveskip=0pt,
}

\title{Combining Example-Based and Rule-Based Program Transformations to Resolve Build Conflicts}

\author{Sheikh Shadab Towqir}
\affiliation{%
  \institution{Virginia Tech}
  \city{Blacksburg}
  \state{Virginia}
  \country{USA}
}
\email{shadabtowqir@vt.edu}

\author{Fei He}
\affiliation{%
  \institution{Tsinghua University}
  \city{Beijing}
  \country{China}}
\email{hefei@mail.tsinghua.edu.cn}

\author{Todd Mytkowicz}
\affiliation{%
  \institution{Google LLC}
  \city{Seattle}
  \country{USA}
}
\email{toddmytkowicz@google.com}

\author{Na Meng}
\affiliation{%
  \institution{Virginia Tech}
  \city{Blacksburg}
  \state{Virginia}
  \country{USA}
}
\email{nm8247@vt.edu}

\begin{abstract}
Merge conflicts often arise when developers integrate changes from different software branches.
The conflicts can result from overlapping edits in programs (i.e., \emph{textual conflicts}), or cause build and test errors (i.e., \emph{build} and \emph{test} conflicts). 
They degrade software quality and hinder programmer productivity.  
While several tools detect \emph{build conflicts}, few offer meaningful support for resolving them. 
To overcome limitations of existing tools, 
we introduce \toolb (\underline{BU}ild \underline{CO}nflict \underline{R}esolver), a new conflict resolver.  
\toolb first detects conflicts by comparing three versions related to a merging scenario: base $b$, left $l$, and right $r$. 
To resolve conflicts, it employs two complementary strategies: 
example-based transformation (\toolba) and rule-based transformation (\toolbb). 
\toolbb applies predefined rules to resolve conflicts in frequently suggested or conventional ways. 
\toolba mines branch versions ($l$ and $r$) for exemplar edits applied to fix related build errors. 
From these examples, it infers and generalizes program transformation patterns 
to resolve conflicts in project-specific or unconventional ways.



We evaluated \toolb on \totalcon real-world build conflicts spanning \totaltype distinct conflict types. \toolb
generated at least one solution for 65 cases, and correctly resolved \totalResolved conflicts.
We observed  
that this hybrid approach---combining context-aware, example-based learning with structured, rule-based resolution---can effectively help resolve conflicts. 
Our research sheds light on future directions of more intelligent and automated merge tools.




\end{abstract}

\ccsdesc[500]{Software and its engineering~Software maintenance tools}
\ccsdesc[300]{Software and its engineering~Maintaining software}
\ccsdesc[300]{Software and its engineering~Software evolution}

\keywords{software merge, build conflict, static analysis, example-based, rule-based, program transformation, example mining, pattern inference}

\maketitle
\vspace{-.5em}
\section{Introduction}
Version control systems (VCSs) like Git are popularly used in collaborative software development. 
With VCSs, programmers create and work on separate branches for feature addition, bug fixing, or feature improvement. Periodically, they merge branches into a primary branch, 
to integrate the edits applied to distinct branches into one program version. In such a scenario, \textbf{merge conflicts} can happen if the edits from different branches are incompatible.
It is challenging and time-consuming to properly handle conflicts. 
Prior work shows that 
developers often spend hours or days detecting
and resolving conflicts before correctly merging branches~\cite{kasi2013cassandra}.

\begin{figure}[h]
    \centering
    \vspace{-1.em}
    \includegraphics[width=.9\linewidth]{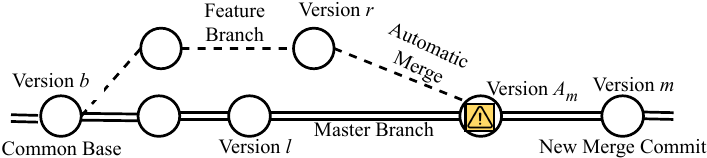}
    \vspace{-1.2em}
    \caption{A merging scenario involves five program versions}\label{fig:merging-scenario}
    \vspace{-1.em}
\end{figure}

\begin{figure*}
\begin{minipage}{.34\linewidth}
    \includegraphics[width=\linewidth]{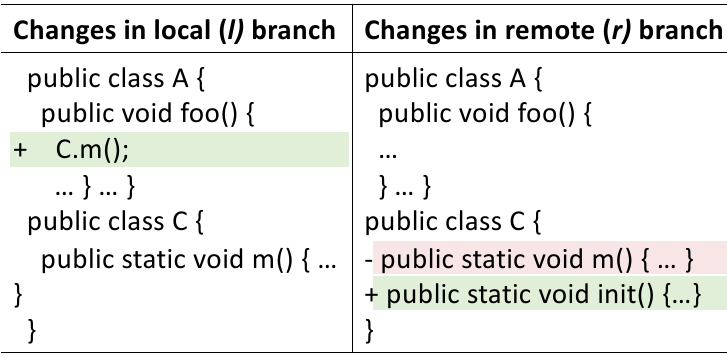}
    \vspace{-2.5em}
    \caption{A build conflict due to the updated def of {\tt m()} by $r$, and added use by $l$}\label{fig:example}
\end{minipage}
\hfill
\begin{minipage}{.64\linewidth}
    \centering
    \includegraphics[width=\linewidth]{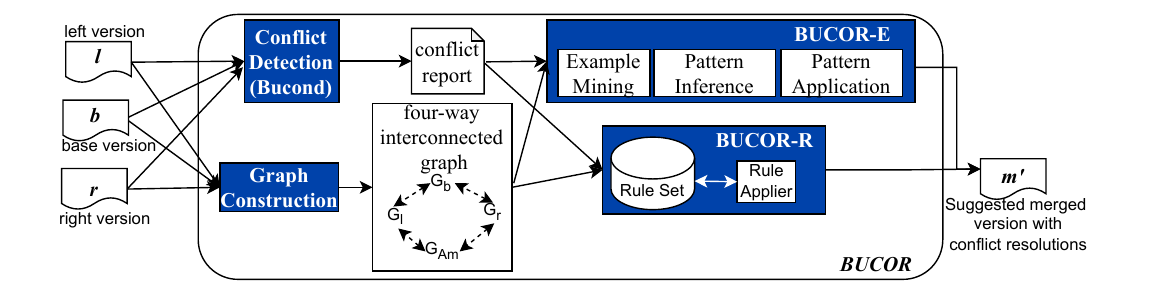}
    \vspace{-2.5em}
    \caption{\toolb leverages Bucond~\cite{towqir2022detecting} to detect build conflicts, and employs two complementary strategies to resolve conflicts}
    \label{fig:overview}
    \vspace{-1em}
\end{minipage}  
\vspace{-1.2em}
\end{figure*}

As illustrated in Figure~\ref{fig:merging-scenario}, a typical merging scenario involves 
five program versions: two \textbf{branch versions} $l$ and $r$ whose edits need to be merged, \textbf{base version} $b$---the common origin of both branches,
the \textbf{automatically merged version} $A_m$ produced by git-merge when it na\"ively integrates branch edits textually, and the \textbf{manually merged version} $m$ that developers create based on $A_m$.
Among the various possible conflicts between $l$ and $r$, \textbf{build conflicts} refer to the incompatible edits whose na\"ive integration triggers build errors in the resulting merged version. 
As shown in Figure~\ref{fig:example}, 
because $l$ adds a call to $m()$ while $r$ renames that method, the co-application of both edits can cause a compilation error of unresolved method reference. Namely, the newly added method call $m()$ is not associated with any defined method. 

Although tools were created to detect build conflicts~\cite{towqir2022detecting,Wuensche2020}, there is rare automatic support to resolve such conflicts.
In particular, 
Gmerge~\cite{zhang2022using} relies on Clang compiler messages to locate conflict-related changes in branches, and applies k-shot learning with GPT-3 to suggest symbol renaming for conflict resolution. 
MrgBldBrkFixer~\cite{Sung2020} starts with a na\"ively merged version $A_m$, looks for build errors due to failed resolution of symbols, identifies symbol renaming changes in branches-to-merge, extracts related patches in either branch, and similarly applies those patches to fix build conflicts. 
These tools suffer from two limitations:

\begin{enumerate}
    \item They only handle conflicts related to symbol renaming (Figure~\ref{fig:example}), but fail to resolve many other conflicts, such as those caused by class-hierarchy change, method addition/deletion, and parameter addition/deletion.   
    \item They suggest simple edits of symbol renaming to adapt \emph{use}s of renamed entities, but fail to suggest complex edits that systematically change \emph{uses} and surrounding context.
\end{enumerate}

To overcome these limitations, 
we introduce \textbf{\toolb (\underline{BU}ild \underline{CO}nflict \underline{R}esolver)}, 
a novel resolver of build conflicts. 
As shown in Figure~\ref{fig:overview}, given the three program versions ($l$, $b$, $r$) related to a merging scenario, \toolb first applies Bucond~\cite{towqir2022detecting}, a static analysis-based tool, 
to reveal conflicting edits between branches. 
\toolb also adopts git-merge to generate a na\"ively merged version $A_m$, extending Bucond to (1) create a graphical representation for $A_m$ and (2) relate that graph with those of the three input program versions.
Based on the generated conflict report and four-way interconnected graph, \toolb opportunistically applies two complementary strategies: 
\toolba and \toolbb. 

\textbf{Example-based transformation (\toolba):}
Prior work shows that many 
build conflicts are caused by mismatches between revised or removed \emph{def} (short for ``definition''), and newly introduced \emph{use} of the same program entities (e.g., classes and methods)~\cite{Shen2022,da2022build}. Inspired by this insight,  \toolba identifies the \emph{def}-change responsible for each reported conflict, locates the owner branch, and mines that branch for 
any exemplar edit applied to adjust corresponding \emph{use}s. It then infers a program transformation pattern from that edit. 
For each \emph{use}-addition responsible for the reported conflict, \toolba tentatively establishes context matching between the pattern and code in $A_m$. 
If a full or partial match is found, 
\toolba customizes and applies the entire or partial pattern for conflict resolution.

\textbf{Rule-based transformation (\toolbb):}
Prior studies reveal patterns frequently applied to resolve certain kinds of build conflicts~\cite{Wuensche2020,Sung2020}. 
Following those studies, we defined and implemented 16 resolution rules/patterns in \toolbb. 
Given a conflict, \toolbb searches its rule set for an  applicable rule; when a rule is found, \toolbb customizes that rule for edit application.

To evaluate the tool effectiveness, we applied \toolb to \totalcon real-world build conflicts in 30 open-source projects. \toolb correctly resolved \totalResolved conflicts. \toolba and \toolbb separately generated resolutions for 28 and 51 conflicts; 21 and 20 of those  resolutions separately match developers' resolutions recorded in version history. All these numbers demonstrate \toolb's effectiveness in handling conflicts. To sum up, we made the following contributions:



\begin{itemize}
    \item We created \toolb, a new static analysis-based resolver of build conflicts, to novelly combine example-based transformation with rule-based transformation. 
    \item We explored \toolba, an advanced example-based approach of conflict resolution. 
    Different from existing tools, it applies program dependency analysis to derive resolution patterns from exemplar edits, establishes context matching for partial/full pattern customization as well as application, and ranks candidate resolutions to suggest the best one. 
    \item We explored \toolbb, a rule-based approach to resolve frequently occurred conflicts in 16 conventional ways. No prior work implements such an approach. 
    \item We systematically evaluated \toolb with a dataset of \totalcon conflicts that span 21 types, and observed novel phenomena. 
\end{itemize}
At \url{https://github.com/shadabtowqir/BuCoR}, we open-sourced our program and data.
\begin{table*}
\footnotesize
\caption{A motivating example of build conflict, whose resolution requires context-specific edits more than symbol renaming
}\label{tab:example}
\vspace{-1.7em}
\centering
\begin{tabular}{p{8.6cm}|p{8.5cm}}
\toprule
\multicolumn{2}{c}{\textbf{Edits from the branches-to-merge}}\\ \toprule
\textbf{(a) Conflicting edits between branches}  & \textbf{(b) \toolb mines branch edits for exemplar edit E to adapt code}\\ \hline
\begin{tabular}{p{8.5cm}}
 Changes in $\boldsymbol{l}$ (responsible def updates):      \\ 
 In {\tt TypeSerializerConfig.java}, names of the file, Java class, and constructor are all updated to {\tt SerializerConfig}.\\
 \\
 Changes in $\boldsymbol{r}$ (responsible use introduction): \\
 In a newly added file {\tt XmlClientConfigBuilder.java}, a method is defined to access {\tt TypeSerializerConfig}\\
\vspace{-.5em}
\begin{lstlisting}
+ private void handleSerializers(Node node, SerializationConfig serializationConfig) {
+  ...
+  if ("type-serializer".equals(name)) {
+   TypeSerializerConfig typeSerializerConfig = new TypeSerializerConfig();
+   typeSerializerConfig.setClassName(value);
+   final String typeClassName = getAttribute(child, "type-class"); 
+   typeSerializerConfig.setTypeClassName(typeClassName);
+   serializationConfig.addTypeSerializer(typeSerializerConfig);
+  } ... }
\end{lstlisting} 
\end{tabular}
& 
\begin{tabular}{p{8.5cm}}
Edit example in $\boldsymbol{l}$ to adapt usage of {\tt TypeSerializerConfig}: \\
\vspace{-.5em}
\begin{lstlisting}
private void handleSerializers(Node node, ...) {
  ...
- if("type-serializer".equals(name)){
-  TypeSerializerConfig typeSerializerConfig = new TypeSerializerConfig();
-  typeSerializerConfig.setClassName(value);
+ if("serializer".equals(name)){
+  SerializerConfig serializerConfig = new SerializerConfig();
+  serializerConfig.setClassName(value);
   final String typeClassName=retrieveAttribute(child, "type-class");
-  typeSerializerConfig.setTypeClassName(typeClassName);
-  serializationConfig.addTypeSerializer(typeSerializerConfig);
+  serializerConfig.setTypeClassName(typeClassName);        
+  serializationConfig.addSerializerConfig(serializerConfig);
\end{lstlisting} 
\end{tabular}
\\ \toprule
\multicolumn{2}{c}{\textbf{Tool-generated edits that are applicable to $A_m$, to resolve conflicts}} \\ \toprule
\textbf{(c) A na\"ive resolution producible by existing tools and \toolbb} & \textbf{(d) A more comprehensive resolution produced by \toolba} \\ \hline 
\vspace{-.5em}
\begin{lstlisting}
 private void handleSerializers(Node node, ...) {
  ...
  if ("type-serializer".equals(name)) { 
(*@\textcolor{red}{- \ \ TypeSerializerConfig typeSerializerConfig = new TypeSerializerConfig();} @*)
(*@\textbf{\textcolor{ForestGreen}{+ \ \ SerializerConfig typeSerializerConfig = new SerializerConfig();}} @*)
   typeSerializerConfig.setClassName(value);
   final String typeClassName=getAttribute(child, "type-class");
   typeSerializerConfig.setTypeClassName(typeClassName);
   serializationConfig.addTypeSerializer(typeSerializerConfig);
  } ... }
\end{lstlisting} 
&
\vspace{-.5em}
\begin{lstlisting}%[language=Java,breaklines=true]
 private void handleSerializers(Node node, ...) {
  ...
(*@\textcolor{red}{- if ("type-serializer".equals(name2)) \{} @*)
(*@\textcolor{red}{- \ TypeSerializerConfig typeSerializerConfig = new TypeSerializerConfig();} @*)
(*@\textcolor{red}{- \ typeSerializerConfig.setClassName(value);} @*)
(*@\textbf{\textcolor{ForestGreen}{+ if ("serializer".equals(name2)) \{ }} @*)
(*@\textbf{\textcolor{ForestGreen}{+ \ SerializerConfig serializerConfig = new SerializerConfig();}} @*)
(*@\textbf{\textcolor{ForestGreen}{+ \ serializerConfig.setClassName(value);}} @*)
  final String typeClassName=getAttribute(child, "type-class");
(*@\textcolor{red}{- \ typeSerializerConfig.setTypeClassName(typeClassName);} @*)
(*@\textcolor{red}{- \ serializationConfig.addTypeSerializer(typeSerializerConfig);} @*)
(*@\textbf{\textcolor{ForestGreen}{+ \ serializerConfig.setTypeClassName(typeClassName);}} @*)
(*@\textbf{\textcolor{ForestGreen}{+ \ serializationConfig.addSerializerConfig(serializerConfig); }} @*)
  } ... }
\end{lstlisting} 
\\ \bottomrule
    \end{tabular}
\vspace{-1.em}
\end{table*}

\vspace{-1.em}
\section{A Motivating Example}
\label{sec:project2_background}

To facilitate discussion, we introduce a running example we crafted based on a real-world project 
hazelcast~\cite{hazelcast-conflict}. 
As shown in Table~\ref{tab:example}(a), 
a merging scenario has $l$ rename a class \codefont{TypeSerializerConfig} and $r$ insert code to use the original class.
The na\"ive integration of these edits
can trigger a build error as the newly introduced \emph{use}s refer to a nonexistent class \emph{def}; thus, a build conflict occurs.

To resolve such conflicts, existing tools update class references by replacing \codefont{TypeSerializerConfig} with \codefont{SerializerConfig} (Table~\ref{tab:example}(c)). However, such a replacement is insufficient, as the context still has variable \codefont{typeSerializerConfig} and literal ``\codefont{type-serializer}'' match the original class. 
Consequently, after existing tools update class references, developers need to manually replace \emph{def}s and \emph{use}s of related variables/literals, to ensure consistent updates and prevent semantic conflicts. When such class \emph{use}s are introduced at multiple places, developers have to go over all places to manually apply those edits again and again, which process is tedious and error-prone. 

To overcome the limitations of existing tools and better help developers, we introduce \toolb---a hybrid approach to combine example-based resolution with rule-based resolution. The example-based resolution \toolba is derived from our insight, on the association between build errors in branches and build conflicts in the merged version. Basically, 
many build errors are caused by mismatches between \emph{def} and \emph{use} of the same program entities.
There is commonality between (1) fixes to build errors on branches, and (2) resolutions to build conflicts in the merged version
~\cite{Shen2022,da2022build}. 
\emph{If on either branch, developers resolved def-use mismatches by applying specialized edits; 
then they are likely to reapply similar edits to resolve conflicts that show the same kind of mismatches
in software merge.} 

For each reported conflict, \toolba locates the responsible \emph{def}-change, and mines the contributing branch for developers' exemplar edit $E$ applied to adjust existing \emph{use}s of the changed entity. 
 As shown in Table~\ref{tab:example}(b), the edit example is extracted from $l$ by \toolb, because that edit adjusts usage of \codefont{TypeSerializerConfig}---the renamed class. The edit  $E$ not only updates references to the old class/constructor, but also revises a related variable \codefont{typeSerializerConfig}, a literal ``\codefont{type-serializer}'', and a method call \codefont{addTypeSerializer(...)}. 
\toolba then derives a transformation pattern $P$ from $E$, by abstracting away irrelevant edit detail and/or program context. 

For the responsible \emph{use}-introduction of each  conflict, \toolba establishes context matching between $P$ and the edit location. If a match is found, \toolba customizes $P$ by creating edit operations with respect to the matched context. \toolb then applies those operations to transform code. As shown in Table~\ref{tab:example}(d),  the context-to-handle is different from the original edit context (see Table~\ref{tab:example}(b)): it uses a different variable (\codefont{name2} vs.~\codefont{name}), and calls a different method (\codefont{getAttribute(...)} vs.~\codefont{retrieveAttribu\-te(...)}). However, \toolb could resolve the conflict by mimicking developers' edits.

In addition to \toolba, our tool also integrates a rule-based transformation approach \toolbb. This is because there are scenarios where $l$ and $r$ do not have exemplar edit $E$
in response to \emph{def}-changes, making \toolba less useful. 
\toolbb can opportunistically suggest resolutions when \toolba does not work, to mitigate the limitation of example-based transformation. 
For the merging scenario described above, \toolb suggests two alternative resolutions for developers to review (see Table~\ref{tab:example}(c)--(d)).

\vspace{-1em}
\section{Approach}
\label{sec:project2_methodology}
As shown in Figure~\ref{fig:overview}, 
\toolb has four components: conflict detection, graph construction, \toolba, and \toolbb. 
This section explains each of them in detail.

\vspace{-.5em}
\subsection{Conflict Detection (Bucond~\cite{towqir2022detecting})}\label{sec:bucond}
Bucond statically analyzes three program versions related to each merging scenario---$b$, $l$, $r$---to identify build conflicts. 
It models each version as a \textbf{program entity graph (PEG)}, which captures defined program entities (e.g., classes and methods) and their relations (e.g., type reference or method calls).
By comparing PEGs, Bucond extracts entity-level edits in $l$ and $r$. 
It then matches these edits with 57 predefined patterns of conflicting edits.
For instance, one pattern checks when a method is renamed in one branch, whether the other branch adds any call to the original method. 
Bucond 
reports a conflict if any match is found.  
Unlike compiler-based tools, Bucond does not try to generate or build the  na\"ively merged version $A_m$. It can detect conflicts even if $A_m$ is uncompilable, and pinpoint the specific edits responsible---which compilers cannot do.

For implementation, Bucond uses JavaParser~\cite{jp} to parse source code, and adopts JGraphT~\cite{jgrapht} to construct and analyze PEGs.




\vspace{-.5em}
\subsection{Graph Construction}
\label{sec:project2_graph_construction}
To facilitate users' conflict comprehension and our tool's resolution placement, we extended Bucond to construct a \textbf{four-way interconnected graph}. Our extension involves two parts: (1) generating a merged version $A_m$ and (2) comparing its PEG with those of given program versions.
As a first step, \toolb uses the widely used tool
git-merge~\cite{git-merge} to create a na\"ively merged version $A_m$.  
While git-merge can detect textual conflicts---cases where multiple branches apply divergent edits to the same code fragment, it does not detect or resolve build conflicts.
The $A_m$ it produced offers a program context for which \toolb later suggests resolution edits.

\begin{figure}
   \centering
    \includegraphics[width=.9\linewidth]{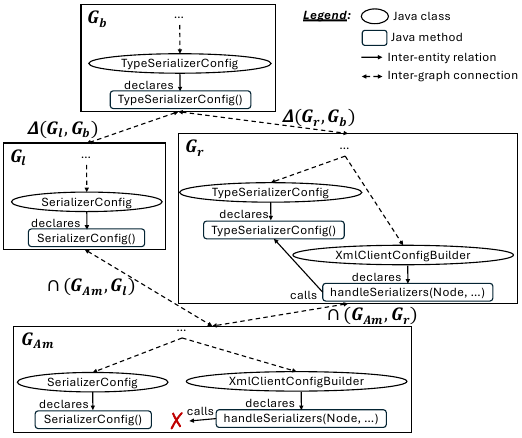}
    \vspace{-.5cm}
    \caption{Four-way interconnected graph for the motivating example in Table~\ref{tab:example}}
        \label{fig:node_graph1}
    \vspace{-1.5em}
\end{figure}

In Step 2, 
to relate $A_m$ with $l$ and $r$, \toolb also creates a PEG for $A_m$, and matches this graph against the PEGs of other versions. As shown in Figure~\ref{fig:node_graph1}, we denote the PEGs as $G_b$, $G_l$, $G_r$, $G_{Am}$, corresponding to the base, left, right, and na\"ively merged versions. 
The differences between $G_l$ and $G_b$, and between $G_r$ and $G_b$, are represented as 
$\Delta(G_l, G_b)$ and $\Delta(G_r, G_b)$; they capture entity-level edits like entity addition/deletion/update, and relation addition/deletion. They are computed by Bucond. 


We use $\cap(G_{Am}, G_l)$ and $\cap(G_{Am}, G_r)$, to separately denote intersections between $G_{Am}$ and $G_l$, and between $G_{Am}$ and $G_r$. They represent structural commonality between versions, serving as anchors to align program context as well as edits between $A_m$, $l$, and $r$. 
\toolb recognizes such commonality using the graph-matching algorithm from Bucond. Intuitively, given two graphs under comparison, the algorithm first identifies exact node matches based on entity types and fully qualified names (FQNs). For nodes that remain unmatched, it ambiguously matches nodes based on the similarity of their internal code implementation and 
surrounding context. 
The identified common elements across program versions enable  \toolb to accurately position edits during conflict resolution. 


\vspace{-.5em}
\subsection{Example-based Transformation (\toolba)}
\label{sec:project2_example_based}
To resolve a conflict, \toolba operates in three phases. 
It first mines branches for edit example(s), infers a transformation pattern from each example, and then customizes as well as applies those patterns to revise $A_m$.



\subsubsection{Phase I. Example Mining}
\label{sec:project2_motivating_example}
Given a conflict, \toolba identifies the responsible \emph{def}-change, mines the branch contributing that change, and looks for any exemplar edit which adapts entity \emph{use}s and surrounding code for that change. 
In our motivating example, because $l$ renames class \codefont{TypeSerializerConfig} and its constructor, it is the contributor branch of \emph{def}-change. By traversing  entity-level edits of $l$---$\Delta(G_l, G_b)$, \toolba locates changes of the corresponding class/constructor references. 
Namely, if an entity $\boldsymbol{e}$ uses \codefont{TypeSerializerConfig} in $b$ but uses \codefont{SerializerConfig} in $l$, then the entity $e$ contains an edit example. Certainly, if multiple entities adapt their usage to the same \emph{def}-change, \toolba 
considers all these entities to have relevant examples.

\begin{figure*}
    \centering
    \includegraphics[trim={0cm 0cm 0cm 0cm},clip,width=1\textwidth]{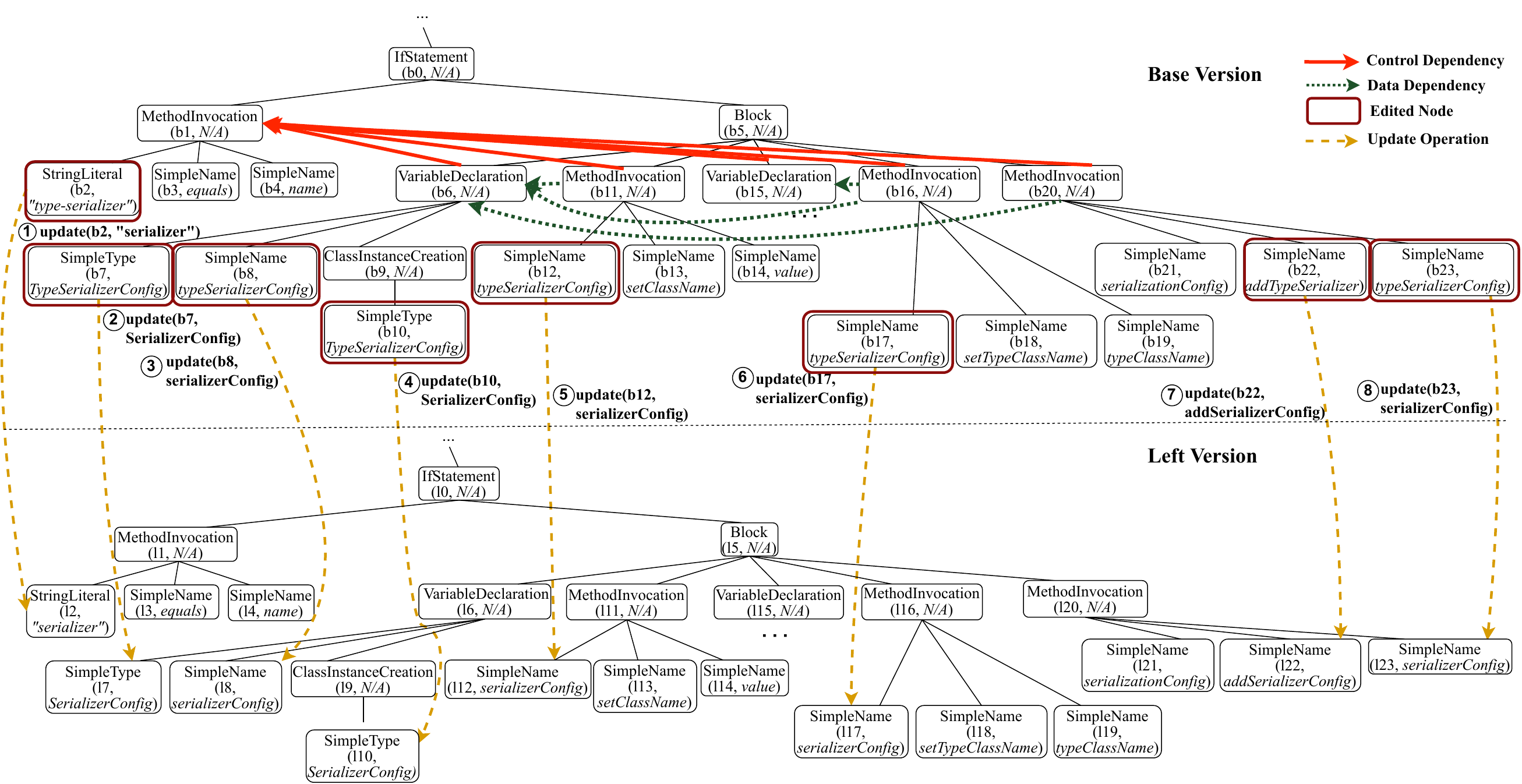}
    \vspace{-2.5em}
    \caption{From the edit example shown in Table~\ref{tab:example}(b), eight AST edit operations are extracted (see \textcircled{1}--\textcircled{8})}
    \label{fig:ast_differencing}\vspace{-1.5em}
\end{figure*}

To extract and represent each edit example, \toolba applies an off-the-shelf syntactic differencing tool GumTree~\cite{Falleri14} to the base and branch versions of $e$, to generate an edit script of Abstract Syntax Tree (AST). The script may have four kinds of operations: 

\begin{itemize}
    \item \codefont{update($t$, $v_n$)}: To replace the old value of node $t$ with the new value $v_n$.
    \item \codefont{add($t$, $t_p$, $i$, $l$, $v$)}: To add a new node $t$ to the AST, as the $i^{th}$ child of node $t_p$. Here, $l$ and $v$ separately specify $t$'s entity type (e.g., method invocation) and its value (e.g., the statement string).
    \item \codefont{delete($t$)}: To remove a node $t$ from the AST.
    \item \codefont{move($t$, $t_p$, $i$)}: To move a node $t$ and its subtree to the $i^{th}$ child of node $t_p$.
\end{itemize}
For our motivating example, the script extracted from the located exemplar edit includes eight operations: each operation updates a node to replace a string literal, type usage, constructor/method usage, or variable usage (see Figure~\ref{fig:ast_differencing}).

\subsubsection{Phase II. Pattern Inference}
In each example $E$, not every edit operation was applied to adapt \emph{use}s for a given \emph{def}-change, and not all unchanged code is relevant to those adaptive changes. To infer a minimal transformation pattern $P$, \toolba takes two steps: edit refinement and context refinement. 

\paragraph{Step 1 of Phase II: Edit Refinement.} 
\toolba extracts edit operations related to \emph{use}s for a given \emph{def}-change, but abstracts away  irrelevant operations. 
Specifically, if a  \emph{def}-change revises an existing field or method, \toolba considers all accesses of that field/method as \emph{use}s. If a \emph{def}-change revises a class, \toolba considers (1) all accesses to that class and (2) all accesses to the variables instantiated with that class as \emph{use}s. 
For our motivating example, as shown in Figure~\ref{fig:ast_differencing}, edit operation \textcircled{2} is relevant as it updates usage of the changed class \codefont{TypeSerializerConfig}; \textcircled{4} is relevant as it updates usage of the changed constructor \codefont{TypeSerializerConfig}; operations \textcircled{3}, \textcircled{5}, \textcircled{6}, and \textcircled{8} are relevant as they all update usage of variable \codefont{typeSerializerConfig}---an instance of \codefont{TypeSerializerConfig}.

Starting with the initial set of
\emph{use}-related operations ($E_0$), \toolba further includes operations 
that occur within the same statement(s) as any operation in $E_0$.
For instance, operation \textcircled{7} gets incorporated because it appears in the same statement as \textcircled{8}. 
Our rationale is that when two edit operations are applied to the same statement, they are very likely to be  semantically related. 
To conservatively preserve the completeness of extracted edits, we include all operations co-applied with \emph{use}-related operations in the same statements. We denote the expanded operation set with $E_1$.

As shown in Figure~\ref{fig:ast_differencing}, \toolba locates edited nodes for the entire edit script in base $b$, identifies the subset of edited nodes covered by $E_1$, and conducts static analysis to find other edited nodes directly depended on by the subset. \toolba extracts two kinds of statement-level dependencies: 


\begin{itemize}
    \item \textbf{Control Dependency:} $x$ depends on  $y$, if whether or not $x$ is executed depends on the execution outcome of $y$.
    \item \textbf{Data Dependency:} $x$ depends on $y$ if $x$ uses a variable defined by $y$.
\end{itemize} 
Such dependencies are essential to reveal changes that constrain any \emph{use}-change. Namely,  without the operations on which $E_1$ depends, 
$E_1$ can wrongly modify semantics and introduce errors like accessing undefined variables.
\toolba adopts WALA~\cite{wala-repo}, a widely used program analysis tool, to conduct dependency analysis. By extracting \emph{use}-related operations and any operations they depend on, \toolba ensures to abstract away non-essential co-applied edit. 
For our motivating example, \textcircled{1} also gets included, as the \codefont{if}-condition controls whether or not edited nodes in $E_1$ get executed. 

\paragraph{Step 2 of Phase II: Context Refinement.} \toolba extracts the minimal subtree that covers all refined operations, 
 to abstract away non-essential context and derive a conflict-resolution pattern.
Listing~\ref{lst:patch} shows the pattern \toolba infers from Table~\ref{tab:example}(b). Compared with the original example, this pattern is more concise. It has all operations necessary to adapt the usage of an updated class and its constructor. Meanwhile, it removes surrounding program context of the \codefont{if}-statement, and an unchanged statement in that structure: \codefont{final String typeClassName=getAttribute(child, "type-class");}. 

\lstset{frame=tb}
\begin{lstlisting}[label=lst:patch,caption=A pattern inferred from the mined edit example]
- if("type-serializer".equals(name)){
-  TypeSerializerConfig typeSerializerConfig = new TypeSerializerConfig();
-  typeSerializerConfig.setClassName(value);
-  typeSerializerConfig.setTypeClassName(typeClassName);
-  serializationConfig.addTypeSerializer(typeSerializerConfig);
+ if("serializer".equals(name)){
+  SerializerConfig serializerConfig = new SerializerConfig();
+  serializerConfig.setClassName(value);
+  serializerConfig.setTypeClassName(typeClassName);      
+  serializationConfig.addSerializerConfig(serializerConfig);
\end{lstlisting} 

\subsubsection{Phase III: Pattern Application}
To resolve conflicts, \toolba enumerates patterns  inferred for each conflict-responsible entity \emph{use}, tentatively
matching each pattern $P$ with the program context  
to see whether the pattern can be customized and applied. This phase has two steps: 
context matching and AST rewriting.  

\begin{algorithm}
    \caption{The context-matching algorithm}\label{alg:context-match}
    \LinesNumbered
    \small
    \KwIn{Pattern $P$, method-to-edit $M$, changed entity $e$}
    \KwOut{Node matches between the ASTs: $W$}
    \tcc{Find a match for the last usage of $e$}
    Locate a statement $s_p \in P$, which has the last usage of $e$ \\
    Search among $M$ for a statement $s_m$ that best matches $s_p$\\
    \If{match is found}{
    $W.add(s_p, s_m, matchingScore)$\\
    \tcc{Use the matched nodes as anchors to guide tentative matches for siblings}
    \ForEach{sibling statement $l_p$ coming before $s_p$}{
        \ForEach{sibling statement $l_m$ coming before $s_m$}{
            Search for a best match in a backward manner\\
            \If{a best match is found}{
                $W.add(l_p, l_m, bestScore)$
            }
        }
    }
    \ForEach{sibling statement $l_p$ coming after $s_p$}{
        \ForEach{sibling statement $l_m$ coming after $s_m$}{
            Search for a best match in a forward manner\\
            \If{a best match is found}{
                $W.add(l_p, l_m, bestScore)$
            }
        }
    }
    \tcc{Compare parent nodes of $s_p$ and $s_m$}
    Locate the parent node of $s_p$, $p_p\in P$, if there is any\\
    Locate the parent node of $s_m$, $p_m \in M$, if there is any\\
    Check whether $p_m$ matches $p_p$\\
    }
    Repeat 1.3--1.17 by treating $p_p$ as $s_p$ and treating $p_m$ as $s_m$, until no more parent can be found or the parent match fails 
\end{algorithm}

\paragraph{Step 1 of Phase III: Context Matching.} 
Because Phase II 
applies backward dependency analysis to \emph{use}-related edited nodes to infer $P$, the resulting contextual AST typically includes critical nodes (i.e., edited nodes using entity $e$) as leaf nodes, and noncritical ones (i.e., unchanged nodes or edited nodes not using $e$) as inner nodes. 
Our goal is to adapt usage of $e$. Thus, to decide whether an inferred pattern is applicable, we prioritize the match search for critical nodes. Namely, if critical nodes do not match, a pattern is inapplicable even if noncritical nodes match. 
If critical nodes match, a pattern is at least partially applicable even if higher-level noncritical ones do not match. 
We designed an algorithm
to perform \emph{use}-centric, bottom-up context matching.

As shown in Algorithm~\ref{alg:context-match}, 
given a pattern $P$, a method $M$ whose entity \emph{use} causes a conflict, and the changed entity $e$, \toolba 
performs best-effort bottom-up node matching. 
It first locates in $P$ a statement $s_p$ that contains the last usage of $e$, and searches $M$ for a statement $s_m$ best matching $s_p$ (lines 1.1--1.2). 

To compare statements, we check their AST node types and values to calculate a matching score as below: 
\begin{itemize}
    \item \textbf{Type Match:} Two nodes, $x$ and $y$,  have a type match if their node types are identical.  
    Such a match increments the initial matching score (i.e., 0) by 1.
   Type match enables patterns to get applied to the context 
   (1) distinct from the original edit example, but (2) preserving the structural consistency. 
    \item \textbf{Value Match:} Two nodes, $x$ and $y$, have a value match if their strings have a 3-gram similarity score greater than 0.618; 
    the similarity score is added to the total matching score. 
   The string of a simple statement (e.g., \codefont{MethodInvocation}) includes all content of that statement; the string of a complex statement (e.g., loop) only includes the header.  
    This threshold-based approach allows flexible matching across similar code, and increases the 
    applicability of inferred patterns.
\end{itemize}
Particularly, if $x$ and $y$ are identical, they have a type match and the 3-gram similarity is 1, so the overall matching score is 2. 

Once a best match of $s_p$ is found (with score $>$ 1.618), \toolba treats the pair $(s_p, s_m)$ as anchor points to guide sibling match (lines 1.5--1.14). 
Preceding siblings are matched in reverse order, starting from the nearest sibling, while following siblings are matched in forward order. 
The 1.618 threshold reflects the expectation of both a type match and a value similarity of at least 0.618. Here, 0.618 is the golden ratio used in prior work~\cite{Shen2019} to yield robust matching. 

After matching all siblings, \toolba attempts to match the parent nodes of $s_p$ and $s_m$. If this match succeeds, \toolba repeats  1.3--1.17 in Algorithm~\ref{alg:context-match}, using the newly matched blocks as anchors for further match. The procedure continues until 
no further parent matches are possible. 
In the end, the algorithm 
 outputs
best matches of nodes, and their matching scores.

If multiple patterns inferred from Phase II apply to method $M$, 
\toolba ranks them with two metrics derived from Algorithm~\ref{alg:context-match}: 

\begin{itemize}
    \item \textbf{Sum of Matching Scores ($\Sigma_m$):} Adding up the scores of all best matches. 
    \item \textbf{Count of Exact Matches ($C_m$):} Counting the number of exact matches (i.e., matching score = 2).
\end{itemize}
We believe that the greater the contextual similarity between a candidate pattern $P_i$ and $M$, the easier it is to apply that pattern and the more likely the resulting version is correct. Thus, among all applicable patterns, we select the one with the highest $\Sigma_m$. Ties are broken using $C_m$ to favor closer semantic matches.

In our motivating example, the method-to-edit (Table~\ref{tab:example}(c)) shares a highly similar context with the inferred pattern (Listing~\ref{lst:patch}): $\Sigma_m=9.94$ and $C_m=4$. 
All five edited statements in the pattern sequentially find best matches within the method: with four exact matches and a highly similar \codefont{if}-condition. 


\paragraph{Step 2 of Phase III: AST Rewriting.}
To apply the selected pattern to $M$, 
\toolba manipulates the AST of $M$ based on context-matching results, and pretty-prints the updated AST to suggest a resolved version to developers. Although \toolba matches context using statement-level similarity calculation, it can manipulate ASTs at a finer granularity (e.g., updating a symbol). This is because the  edit operations inferred from examples by GumTree are finer-grained; reapplying these operations in new context minimizes the modification and helps prevent unwanted changes. 
Furthermore, if the selected pattern only has partial context to match $M$, only the edit operations corresponding to the matched part get applied; the remaining ones corresponding to unmatched part are not applied. 
In this way, \toolba  resolves conflicts with the best effort, to maximize its applicability and conflict-resolution capability.


\subsection{Rule-based Transformation (\toolbb)}
\label{sec:project2_rule_based}

\toolbb resolves conflicts via rule-based program transformations. To define practical resolution rules or patterns, we manually inspected the rules and data observed by prior empirical studies~\cite{Sung2020,da2022build},  
and expanded those rule sets to generalize rules across similar but different conflict types. Table~\ref{tab:res_strategies} shows the resulting 16 conventional rules we implemented in \toolbb; each rule resolves one type of frequent conflicts.
Among the 16 rules, only 6 rules involve symbol renaming (C1, C6, C7, C11, C13, C15). As prior work only resolves conflicts caused by symbol renaming, they cannot handle 10 of the conflict types \toolbb focuses on. 
Only 8 of the 16 rules are clearly described by prior work: C1, C3, C5--C6, C8, C14--C16. We derived the remaining rules by generalizing across related conflict types.

\begin{table}[t]
\centering
\caption{The 16 resolution rules used in \toolbb}
\label{tab:res_strategies}
\vspace{-1.5em}
\footnotesize

\begin{tabular}{p{.4cm}|p{3.5cm}|p{3.5cm}}
\toprule
\textbf{Idx} & \textbf{Conflict Type} & \textbf{Resolution Pattern} \\ 
\toprule
C1 & Class: rename def vs.~add use & Update the added use 
\\ \hline
C2 & Class: add method def in super vs.~add sub class & Update method def in sub to match super\\ \hline
C3 & Class: change a method’s parameter list in super vs.~add sub class & Update method def in sub to match super \\ \hline
C4 & Class: change a method’s return type in super vs.~add sub class & Update method def in sub to match super \\ \hline
C5 & Import: remove def vs.~add use & Re-add the def (i.e., add back the removed entity import)\\ \hline
C6 & Package: rename def vs.~add use & Update the added use (i.e., update import with the new package name) \\ \hline
C7 & Interface: rename def vs.~add use & Update the added use\\ \hline
C8 & Interface: add method def in super vs.~add class to implement the super & Add/update method def in class to override the new method in super \\ \hline
C9 & Interface: change a method’s parameter list in super vs.~add class to implement the super & Update parameter list in class to match super \\ \hline
C10 & Interface: remove method def in super vs.~add class to implement the super & Remove method def in class to match super \\ \hline
C11 & Interface: rename a method def in super vs.~add class to implement the super & Rename method def in class to match super \\ \hline
C12 & Interface: change a class to implement the interface vs.~change a method's return type in the class & Update return type of method in class to match super \\ \hline
C13 & Field: rename def vs.~add use & Update the added use
\\ \hline
C14 & Field: add def vs.~add def & Remove redundant field definition \\ \hline
C15 & Method: rename def vs.~add use & Update the added use\\ \hline
C16 & Method: add def vs.~add def & Remove redundant method definition \\
\bottomrule
\end{tabular}
\vspace{-2em}
\end{table}

\begin{table}
\centering
\footnotesize
\caption{\totaltype conflict types of the 88 conflicts in our dataset}\label{tab:distributioneval}
\vspace{-2em}
\begin{tabular}{p{.4cm} |p{6.5cm} |R{.5cm} }
\toprule
\textbf{Idx}&\textbf{Conflict Type} & \textbf{\# of Conflicts} \\ 
\toprule
C1&Class: rename def vs. add use & 7 \\ \hline
C2&Class: add method def in super vs.~add sub class & 2 \\ \hline
C3&Class: change a method's parameter list in super vs.~add sub class & 2 \\ \hline
C4&Class: change a method's return type in super vs.~add sub class & 1 \\ \hline
C5&Import: remove def vs.~add use & 5 \\ \hline
C6&Package: rename def vs.~add use & 2 \\ \hline
C8&Interface: add method def in super vs.~add class to implement the super & 6 \\ \hline
C9&Interface: change a method's parameter list in super vs.~add class to implement the super & 1 \\ \hline
C10&Interface: remove method def in super vs.~add class to implement the super & 1 \\ \hline
C11&Interface: rename a method in super vs.~add class to implement the super & 2 \\ \hline
C12&Interface: change a class to implement the super vs.~change a method's return type in the class & 9 \\ \hline
C14&Field: add def vs. add def & 3 \\ \hline
C15&Method: rename def vs. add use & 8 \\ \hline
C16&Method: add def vs. add def &  2\\ \toprule

C17&Class: remove def vs.~add use & 10 \\ \hline
C18&Constructor: change the parameter list vs.~add use & 5 \\ \hline
C19&Field: change a field's type vs.~add use & 1 \\ \hline
C20&Field: remove def vs.~add use & 8 \\ \hline
C21&Method: change the parameter list vs.~add use & 4 \\ \hline
C22&Method: change the return type vs.~add use & 1 \\ \hline
C23&Method: remove def vs.~add use & 8 \\ \bottomrule
\end{tabular}
\vspace{-2em}
\end{table} 

\toolbb has a Rule Applier to opportunistically resolve conflicts based on predefined rules. Namely, for each reported conflict, \toolbb tentatively matches the conflict against documented conflict types. If a match is found, Rule Applier applies helper functions to the responsible entity's \emph{def} or \emph{use}, to rewrite the AST nodes.  

\begin{table}
    \caption{The experiment result of \toolb}
    \label{tab:result}
    \vspace{-1.5em}
    \footnotesize
    \centering
    \begin{tabular}{l|R{1.5cm}R{1.3cm}rr}
    \toprule
    \textbf{}&\textbf{\# of Resolutions Generated}&\textbf{\# of Correct Resolutions}&\textbf{Coverage (C)} & \textbf{Accuracy (A)} \\ \toprule
    \toolba & 28 &21  & 32\% (28/88) &75\% (21/28) \\ 
    \toolbb & 51 &20 &58\% (51/88)&39\% (20/51)\\ \
    \toolb & 79 &41 &74\% (65/88)&52\% (41/79)\\ \bottomrule
    \end{tabular}
    \vspace{-2em}
\end{table}

\vspace{-1em}
\section{Experiment}
\label{sec:project2_results}
To evaluate \toolb, we defined three research questions (RQs):

\begin{itemize}
\item\textbf{RQ1:} How often can \toolb generate resolutions? 
\item\textbf{RQ2:} What percentage of generated solutions are correct? 
\item\textbf{RQ3:} What is the effectiveness comparison between \toolba and \toolbb? 
\end{itemize}
We conducted the experiment 
on a Windows machine with AMD Ryzen 9 8945HS @4.00GHz and 16 GB memory.
We did not empirically compare \toolb with  Gmerge~\cite{zhang2022using} or MrgBldBrkFixer~\cite{Sung2020}, as they are close-sourced tools targeting C/C++ programs while \toolb focuses on Java code.  
The following subsections describe our dataset, evaluation metrics, and results. 

\vspace{-.5em}
\subsection{Dataset}
We constructed our evaluation dataset by reusing the datasets of prior work~\cite{towqir2022detecting,da2022build} with our best effort. Given a merging scenario with at least one known build conflict, we decided whether it is reusable based on the following criteria:

\begin{itemize}
\item[a)] Both $l$ and $r$ build successfully. 
\item[b)] The automatically merged version $A_m$ output by git-merge does not contain any textual conflict.
\item[c)] The build conflict is detectable by Bucond---\toolb's conflict detection module.
\end{itemize}
We ended up with a dataset of 88 build conflicts from 55 merging scenarios, which were mined from in total 30 open-source Java repositories. Most of the cases we filtered out do not satisfy a) or b).
As shown in Table~\ref{tab:distributioneval}, the 88 conflicts belong to 21 types: 14 of the types overlap with those mentioned in \toolbb (see Table~\ref{tab:res_strategies}), the other types (C17--C23) are not resolvable by any of the predefined rules. 
For each conflict, we retrieved and inspected 
developers' merged version $m$, 
 using it as the ground truth of conflict resolution. 

\vspace{-.5em}
\subsection{Metrics}

We defined two evaluation metrics: 

\emph{\textbf{Coverage (C)}} measures among all known conflicts, how many have at least one resolution generated by the resolver:
\begin{equation*}
C = \dfrac{ \text{\# of conflicts with at least one resolution generated}}{\text{Total \# of known conflicts}}
\end{equation*}

\emph{\textbf{Accuracy (A)}} measures among all suggested resolutions, how many of them are correct:

\begin{equation*}
A = \dfrac{ \text{\# of correct resolutions}}{\text{Total \# of generated resolutions}} 
\end{equation*}
A resolution is correct if 
the resolved version is semantically equivalent to the ground truth---developers' resolution recorded in $m$. Two authors independently inspected \toolb's resolution to examine semantic alignment, and had a discussion whenever they disagree.


\vspace{-.5em}
\subsection{RQ1: \toolb Resolution Coverage}
\label{sec:project2_rq1}

As shown in Table~\ref{tab:result}, \toolba and \toolbb generated resolutions for 28 and 51 cases, respectively. 
The coverage metrics they separately achieved are 32\% and 58\%. 
When combining their outputs, \toolb generated at least one resolution for 65 cases, resulting in an overall coverage of 74\% (65/88).
There are 14 cases where both strategies suggested resolutions. Among the remaining, \toolba and \toolbb separately resolved 14 and 37 cases. 
These numbers imply the two resolution strategies to complement each other. 

\begin{lstlisting}[label=lst:example1,
caption=A conflict without edit example in branches]
/* Left Version: PathItem.java adds a method call to setRef(...) */
(*@\textbf{\textcolor{ForestGreen}{
+ pathItemObject.setRef(apiCallback.callbackUrlExpression());}} @*)

/* Right Version: Reader.java renames setRef(...) */
(*@\textcolor{red}{- public void setRef(String ref) \{
} @*)
(*@\textbf{\textcolor{ForestGreen}{
+ public void set\$ref(String \$ref) \{
}} @*)
\end{lstlisting}
\begin{lstlisting}[caption={A conflict whose resolution is unconventional}, label=example2]
/* Left Version: JedisCluster.java removes a field */
(*@\textcolor{red}{- private int timeout;} @*)

/* Right Version: JedisCluster.java adds a reference to field timeout */
(*@\textbf{\textcolor{ForestGreen}{+ public Set<String> spop(final String key, final long count) \{}}@*)
(*@\textbf{\textcolor{ForestGreen}{+ \   return new JedisClusterCommand<Set<String>>(connectionHandler, timeout, maxRedirections) \{}}@*)
(*@\textbf{\textcolor{ForestGreen}{+ \ \   @Override }}@*)
(*@\textbf{\textcolor{ForestGreen}{+ \ \ public Set<String> execute(Jedis connection) \{ }}@*)
(*@\textbf{\textcolor{ForestGreen}{+ \ \ \ return connection.spop(key, count); }}@*)
(*@\textbf{\textcolor{ForestGreen}{+ \ \   \}\ \}.run(key); \} }}@*)
\end{lstlisting}
\begin{lstlisting}[caption={A conflict whose ground-truth resolution partially overlaps with exemplar edits~\cite{hazelcast-example}}, label=example4]
/* Left Version moves a method between classes */
/* ClientMap.java */
(*@\textcolor{red}{-    public final SerializationService getSerializationService() \{...\} } @*)
/* ClientContext.java */
(*@\textbf{\textcolor{ForestGreen}{+    public SerializationService getSerializationService() \{...\} }}@*)  

/* Right Version: ClientMultiMapProxy.java adds calls to the original method */
(*@\textbf{\textcolor{ForestGreen}{+ public boolean put(K key, V value) \{ }} @*)
(*@\textbf{\textcolor{ForestGreen}{+ \ Data keyData = getSerializationService().toData(key);}} @*)
(*@\textbf{\textcolor{ForestGreen}{+ \ Data valueData = getSerializationService().toData(value);}} @*)
(*@\textbf{\textcolor{ForestGreen}{+ \ PutRequest request = new PutRequest(proxyId, keyData, valueData, -1, ThreadUtil.getThreadId());}} @*)
(*@\textbf{\textcolor{ForestGreen}{+ \ Boolean result = invoke(request, keyData);}} @*)
(*@\textbf{\textcolor{ForestGreen}{+ \          return result; }} @*)
(*@\textbf{\textcolor{ForestGreen}{+    \} }} @*) 
\end{lstlisting}
\begin{lstlisting}[caption={A conflict whose ground-truth resolution has no overlap with exemplar edits~\cite{redisson}}, label=example3]
/* Left Version: RedisClient.java adds a call to one of its constructors */
(*@\textbf{\textcolor{ForestGreen}{+ public RedisClient(String host, int port, int connectTimeout, int commandTimeout) \{ }}@*)
(*@\textbf{\textcolor{ForestGreen}{+ \ this(new HashedWheelTimer(), new NioEventLoopGroup(), NioSocketChannel.class, host, port, connectTimeout, commandTimeout); }}@*)
(*@\textbf{\textcolor{ForestGreen}{+ \} }}@*)

/* Right Version: RedisClient.java adds a parameter to one of its constructors */
(*@\textcolor{red}{- public RedisClient(final Timer timer, EventLoopGroup group, Class<? extends SocketChannel> socketChannelClass, String host, int port, int connectTimeout, int commandTimeout) \{ } @*)
(*@\textbf{\textcolor{ForestGreen}{+ public RedisClient(final Timer timer, ExecutorService executor, EventLoopGroup group, Class<? extends SocketChannel> socketChannelClass, String host, int port, int connectTimeout, int commandTimeout) \{ }}@*)
\end{lstlisting}

Particularly, \toolba suggests nothing when no example exists or no exemplar edit is located in branches.
For example, Listing~\ref{lst:example1} shows a conflict from {swagger-core}~\cite{swagger-core}. The conflict is of type C15---\emph{Method: rename def vs.~add use}, where $r$ renames method \emph{setRef(...)} and $l$ adds an invocation to that method. 
Although an intuitive resolution is to update the newly added method call, no edit example in $r$ demonstrates such change.  
On the other hand, we applied GumTree to compare different versions of source code, and derive AST edit scripts to represent code changes. As GumTree cannot detect any changes to \codefont{PackageDeclaration} or \codefont{ImportDeclaration}, \toolba cannot extract such edits to suggest related resolutions.


\toolbb is applicable to many cases where (1) \toolba does not work and (2) the needed resolutions are conventional, such as the case in Listing~\ref{lst:example1}. 
 However, for conflicts which have no conventional resolution method recommended by literature (see C17--C23 in Table~\ref{tab:distributioneval}),  
 \toolbb generates nothing because there is no typical, generally accepted way to handle them. For example, 
 Listing~\ref{example2} shows a conflict from Jedis~\cite{jedis}. This conflict is of C20--\emph{Field: remove def vs.~add use}, where $l$ removes field \codefont{timeout} and $r$ introduces a new reference to \codefont{timeout}. 
Such a conflict may get resolved by adding back the field \emph{def}, replacing the added \emph{use} with something else, or simply removing that \emph{use}; nevertheless, there is no standard solution or typical resolution pattern followed by developers or documented in literature. Therefore, \toolbb did not handle it, while \toolba resolved it in a project-specific way by referring to edit examples. It removed field \emph{use} as below: 

\codefont{return new JedisClusterCommand<Set<String>>(connectionHandler, maxRedirections) \{...\}}


Among the 21 conflict categories in our dataset, \toolb resolved conflicts of 19 categories. It did not resolve any conflict of C19 and C22, as (1)  
\toolba did not find any exemplar edit, and (2)
there is no well-accepted resolution pattern for \toolb to implement.

\noindent\begin{tabular}{|p{8.1cm}|}
	\hline
	\textbf{Finding 1 (Answer to RQ1):} \emph{\toolb generated resolution(s) for 65 of the given 88 conflicts, achieving 74\% coverage. Among the 79 resolutions it produced, 28 and 51 were separately contributed by \toolba and \toolbb.}
	\\
	\hline
\end{tabular}

\vspace{-.5em}
\subsection{RQ2: \toolb Resolution Accuracy}
\label{sec:project2_rq2}

As shown in Table~\ref{tab:result}, 
21 of the 28 resolutions produced by \toolba are correct, leading to 75\% accuracy; 20 of the 51 resolutions output by \toolbb are correct, achieving 39\% accuracy. Overall, \toolb 
obtained 52\% accuracy, having 41 of the 79 generated resolutions to be correct.
These measurements also show the two strategies to complement each other. 
Among the 14 cases where both strategies suggest resolutions, \toolba and \toolbb correctly resolved 13 and 7 cases, respectively; the correct resolutions overlap in 7 cases. 
Thus, \toolb resolved \totalResolved cases correctly.  

Particularly, 
\toolba resolved build conflicts by mimicking developers' project-specific solutions to related build errors.
However, \toolba did not always output correct solutions, because the expected ground-truth resolutions of some conflicts partially overlap with fixes to build errors, or present patterns totally different from those fixes. 
For instance, in Listing~\ref{example4}, $l$ moves method \codefont{getSerializationServ\-ice()} from class \codefont{ClientMap} to \codefont{ClientContext}, while $r$ adds two calls to the original method. \toolba replaces the first method call with  \codefont{getContext().getSeria\-lizationService()}, by following a mined example. 
Although this replacement is correct, \toolba does not replace the second method call as expected since the original example only contains and updates one method call. 



In another scenario (see Listing~\ref{example3}), $r$ updates a constructor's signature by inserting a parameter of type \codefont{ExecutorService}, and $l$ adds a call to the original constructor. \toolba derived a resolution from branch edits, by updating the call to take an extra argument:

\codefont{Executors.newFixedThreadPool(Runtime.getRuntime().availabl\-eProcessors() * 2)}. 

\noindent
Although this edit can resolve the conflict, developers' manual resolution removes the constructor entirely.
In such scenarios,
branch edits do not  reflect developers' preference of conflict resolution. 

\begin{lstlisting}[caption={A conflict for which the well-accepted resolution pattern does not match developers' manual resolution~\cite{fastjson}}, label=example5]
/* Left Version: TypeUtils.java removes an import */ 
(*@\textcolor{red}{- import java.beans.Introspector;}@*)

/* Right Version: TypeUtils.java adds a reference to the imported class */
(*@\textbf{\textcolor{ForestGreen}{+ if (compatibleWithJavaBean)\{ }}@*)
(*@\textbf{\textcolor{ForestGreen}{+ \       propertyName=   Introspector.decapitalize(methodName.substring(3)); }}@*)

\end{lstlisting}


\toolbb suggested incorrect resolutions for three reasons. First, 
its resolution rules
do not always match developers' practices. 
For instance, the conflict in Listing~\ref{example5} has $l$ remove an import and $r$ add a reference to the removed imported class. \toolbb resolved this conflict by adding back the removed import. However, developers implemented a replacement method, to avoid the removed dependency.
Second, for conflicts due to class-hierarchy changes, \toolbb modified subclasses to match changes in super classes, while developers sometimes adapted super classes to subclasses.
Third, when conflict resolution requires implementation of a new method or updates of a method's parameter/return type, \toolbb can only change the method signature, without customizing method implementation accordingly.

The \totalResolved conflicts correctly resolved by \toolb fall into 12 categories, implying that \toolb can properly resolve diverse conflicts.
There are seven conflict categories for which \toolb produced some resolutions but none of them are correct: C2--C4, C8--C9, C12, and C18. Six of the categories (C2--C4, C8--C9, and C12) are about mismatches between super types and sub types; three of the categories are caused by parameter list changes (C3, C9 and C18). 
Among the seven incorrect resolutions output by \toolba, two are partially incorrect, by presenting subsets of the needed changes; five are totally irrelevant. Among the 31 incorrect resolutions output by \toolbb, 15 are partially incorrect, while 16 are irrelevant.


\noindent\begin{tabular}{|p{8.1cm}|}
	\hline
	\textbf{Finding 2 (Answer to RQ2):} \emph{
    \toolb generated in total 41 correct resolution(s), achieving 52\% (41/79) accuracy. 
    \toolba and \toolbb separately contributed 21 and 20 correct resolutions. 
    }
	\\
	\hline
\end{tabular}

\vspace{-.5em}
\subsection{RQ3: \toolba vs.~\toolbb}
\label{sec:project2_rq3}

\toolba and \toolbb are different in two aspects:  

\emph{First, \toolbb has higher coverage (58\% vs.~32\%) by resolving more conflicts; 
\toolba achieves higher accuracy by having more of its resolutions match developers' intents (75\% vs.~39\%).}   
It implies that if developers (1) want to automatically resolve as many conflicts as possible in conventional ways, and (2) have good testing to detect wrong resolutions, they can rely more on \toolbb.
If developers (1) want to automatically resolve  conflicts in unconventional ways but (2) 
have poor support to detect wrong resolution, 
they can rely more on \toolba.
For instance, when both strategies output resolutions (see Table~\ref{tab:example}), it is likely that  \toolba's result is better.

\emph{Second, the two strategies are good at resolving different types of conflicts.} Among the \totalResolved conflicts correctly handled by \toolb, 
7 conflicts are correctly resolved by both strategies.
In addition to that, 
\toolba correctly resolved 14 cases; for 8 of these cases, \toolbb could not suggest anything. The eight cases are related to entity removal (i.e., C17, C20, C23) and parameter-list changes (i.e., C21). 
It means that \toolba is better at resolving conflicts in unconventional ways. Meanwhile, \toolbb correctly resolved 13 cases, for which \toolba suggested nothing. 
These 13 conflicts are mainly related to super-sub mismatches (C10--C11) and duplicated entities (C14, C16). 
It means that \toolbb is better at handling conflicts in conventional ways.

\vspace{-0.5em}
\subsection{Runtime Overhead}

On our dataset, \toolb spent 0.1--53.6 minutes on each of the 55 merging scenarios, with 2.9 minutes as the mean and 0.8 minutes as the median. Among the four components of \toolba, conflict detection and graph construction are the most time-consuming, roughly taking up 55\% and 36\% of the overall runtime. 
The time \toolb spent on each merging scenario is closely related to the number of (1) Java files and (2) program entities. Namely, the more files to parse and the more entities to analyze, the longer \toolb takes. We noticed that \toolb spent the least time 0.1 minutes, when there are 20 Java files and 1,229 program entities under processing. It spent the most time 53.6 minutes when there are 1,472 Java files and 48,299 entities being analyzed.  



\vspace{-1em}
\section{Threats to Validity}
\label{sec:project2_discussion}
\emph{Threats to External Validity.} 
Our evaluation dataset consists of 88 real-world build conflicts, spanning 21 distinct conflict types. However, the program contexts and conflict patterns covered by this dataset may not fully capture the diversity of build conflicts in real world.
\toolbb includes 16 well-defined rules of handling frequently occurring conflicts. Although they cover all the typical conflict-resolution strategies mentioned by prior work~\cite{da2022build,Shen2022}, they may not include all the frequently applied strategies in reality. 
In the future, to enhance our research representativeness, we will (1) expand our dataset to include more merging scenarios, and (2) enlarge the ruleset of \toolbb whenever possible. 

\emph{Threats to Internal Validity.} 
\toolba leverages the threshold 0.618 to decide whether two AST nodes are similar enough to have a value match, because prior work~\cite{Shen2019} shows that this setting led to reasonably good results. However, there can be scenarios where two matching AST nodes have a lower similarity score than 0.618, disabling \toolba to apply edits as expected. 
In the future, we will explore better ways of node matching.

\emph{Threats to Construct Validity.} \toolb leverages Bucond to detect conflicts. If Bucond cannot detect certain build conflicts (e.g., conflicts in build scripts), \toolb cannot resolve those conflicts either. Prior work~\cite{towqir2022detecting} shows that Bucond has a very good coverage of conflict types, and there is a low chance of Bucond to miss real-world conflicts in Java source code. Therefore, the effectiveness of \toolb is not considerably limited by Bucond.


\vspace{-1em}
\section{Related Work}
The related work of our research includes automated software merge, and empirical studies on merge conflicts. 

\vspace{-.5em}
\subsection{Automated Software Merge}
\textbf{Software Merge Tools}~\cite{Apel:2011,Brun:2011,Apel:2012,Guimaraes:2012,Zhu:2018,Shen2019}. 
FSTMerge~\cite{Apel:2011,Cavalcanti17,jFSTMerge} 
parses Java code for ASTs, and matches nodes between branches based on the class or method signatures; it then  integrates the edits inside each matching pair via textual merge. 
{JDime}~\cite{Apel:2012} also parses Java code for ASTs; however, unlike FSTMerge, it merges code purely via structural matching between ASTs and tree manipulation. 
{AutoMerge}~\cite{Zhu:2018} extends JDime, by proposing multiple alternative strategies to resolve conflicts between branches, with each strategy integrating branch edits in a unique way.
IntelliMerge~\cite{Shen2019} presents a graph-based refactoring-aware merging algorithm. 
These tools 
can resolve some textual conflicts, but cannot handle build  conflicts. 
Crystal~\cite{Brun:2011} and WeCode~\cite{Guimaraes:2012} help reveal three kinds of conflicts. Both tools first apply textual merge to create a merged software, and reveal textual conflicts along the way. They then adopt automatic build and testing to find build/test errors,
regarding those errors as indicators of build and test conflicts. However, neither tool pinpoints or resolves build/test conflicts.  

\textbf{Detectors of Build Conflicts}~\cite{Wuensche2020,towqir2022detecting}.
The existing tools both statically analyze branch versions, reason about branch edits, and contrast extracted edits with predefined patterns to detect conflicts. However, neither tool automatically resolves build conflicts. 


\textbf{Detectors of Test Conflicts}~\cite{sousa2018verified,DASILVA2024}.  
{SafeMerge}~\cite{sousa2018verified}
takes in $b$, $l$, $r$, and $m$, for a given merging scenario. It statically infers the relational postconditions of distinct versions to model program semantics. By comparing postconditions, SafeMerge decides whether $m$ is \emph{free of conflicts}, i.e., without introducing new semantics nonexistent in $l$ or $r$. 
SAM~\cite{DASILVA2024} also takes in four program versions related to a merging scenario. It randomly generates tests with EvoSuite, to explore situations where $b$, $l$, $r$ pass a test but $m$ fails that test. 

\textbf{Resolution of Textual Conflicts}~\cite{pan2021can,aldndni2023automatic,aldndni2024,Elias2023,dinella2021deepmerge,Svyatkovskiy22MergeBERT}.  
RPredictor~\cite{aldndni2023automatic,aldndni2024} and MEST\-RE~\cite{Elias2023} train machine learning (ML) models with traditional algorithms (e.g., Random Forest), using features to characterize conflict content, merging scenarios, evolution history, 
developer experience, and/or edits applied by different branches.
Given a conflict, the models then predict 
what resolution strategy to apply (e.g., keep left). 
DeepMerge~\cite{dinella2021deepmerge} and  MergeBERT~\cite{Svyatkovskiy22MergeBERT} employ deep-learning to resolve specified conflicting chunks automatically. 
Pan et al.~\cite{pan2021can} defined a domain-specific language (DSL) to capture repetitive resolution patterns, and proposed a program synthesis approach to learn resolution strategies as DSL programs from example resolutions.


\textbf{Resolvers of Build Conflicts}~\cite{Sung2020,zhang2022using}.  
MrgBldBrkFixer~\cite{Sung2020} compares the ASTs of C++ files, to resolve conflicts related to renaming changes. 
Gmerge~\cite{zhang2022using} applies few-shot learning to GPT-3, to resolve conflicts due to renaming. 
They are analogous to \toolba, because they all infer and apply patterns by analyzing exemplar edits. 
We did not empirically compare \toolb with either tool,  as they are closed-source and target C/C++ conflicts. 

To assess the effectiveness of large language models (LLMs) 
on our dataset, we recently conducted an extra experiment. We prompted OpenAI's o4-mini with the full diffs of (base, left) and (base, right) of each conflicting merging scenario. 
Two authors manually validated o4-mini's outputs. Due to token limit, o4-mini was applicable to only 48 scenarios. It detected 117 conflicts, of which 63 were correct. It correctly resolved 25 of the conflicts---substantially worse than \toolb. 


\toolb is more advanced in three aspects. First, it defines a comprehensive set of 16 resolution rules: in addition to conflicts due to renaming changes, these rules also handle conflicts due to class-hierarchy change and entity addition/deletion. 
Second, it defines sophisticated algorithms to (1) extract resolution-related edits from branches via control- and data- dependency analysis, (2) match patterns with program context in a \emph{use}-centric manner, and (3) derive candidate resolutions to the same conflict by tentatively matching alternative patterns with the same edit context.
Thus, in addition to symbol renaming, it resolves conflicts through systematic code editing to consistently revise program semantics. 
Third, it defines a hybrid approach between example-based and rule-based resolutions, to combine the advantages of both methodologies.

\vspace{-.5em}
\subsection{Empirical Studies of Merge Conflicts}
\textbf{Relationship between Merge Conflicts and Software Maintenance}~\cite{estler2014awareness,ahmed2017empirical,mahmoudi2019refactorings,Vale2023}. 
Estler et al.~\cite{estler2014awareness} surveyed 105 student developers, to study the relationship between awareness (i.e., knowing ``who's changing what'') and merge conflicts. 
Ahmed et al.~\cite{ahmed2017empirical} explored the effect of bad design (code smells) on merge conflicts. 
Mahmoudi et al.~\cite{mahmoudi2019refactorings} studied the relation between merge conflicts and 15 popular refactoring types. 
These researchers showed that conflicts are widespread, although they studied textual conflicts.

\textbf{Characterization of Merge Conflicts}~\cite{ghiotto2018nature,Shen2022,SHEN2024,da2022build}. Ghiotto et al.~\cite{ghiotto2018nature} and Shen et al.~\cite{SHEN2024} characterized textual conflicts from open-source Java projects in terms of number of chunks, size, program constructs involved, their validity (i.e., true vs.~false positives), edit types, file types,  and/or manual resolution strategies. 
However, neither work studies build conflicts. 

Shen et al.~\cite{Shen2022} followed the methodology of Crystal~\cite{Brun:2011} and WeCode~\cite{Guimaraes:2012} to reveal 3 types of conflicts in 208 open-source repositories. They manually inspected in total 538 conflicts to characterize the root causes and developers' resolution strategies of all conflict types.
Da Silva et al.~\cite{da2022build} collected 239 build conflicts to study their root causes and resolution patterns. 
Both studies reported that 
most build conflicts are caused by declarations removed or updated by one branch but referenced by another branch. 
Da Silva et al.~ also observed that conflicts caused by renaming are often resolved by updating the missing reference, whereas removed declarations are often reintroduced. 
Both studies motivated our research, and present the initial datasets for us to use when creating our own evaluation dataset.

\vspace{-1em}
\section{Conclusion}
\label{sec:project2_conclusion}
This paper introduces \toolb, a novel conflict resolver to conduct example-based and rule-based transformation. 
It applies program static analysis to (1) detect conflicts, (2) mine edit examples in branches, (3) derive transformation patterns from mined examples, (4) and apply inferred or predefined patterns to resolve conflicts. Compared with prior work, 
it significantly pushes the boundary of automatic conflict resolution. Our investigation is deeper as
(a) the conflicts we focus on are very diverse; 
(b) the edits we suggest vary a lot depending on the program context and conflict types;
(c) in the scenarios where no standard resolution pattern exists, \toolb can create resolutions by locating, refining, and reusing relevant edits. In the future, we will improve \toolb by (i) expanding the rule set of \toolbb, and (ii) improving the example mining capability of \toolba when no local example is extractable from branches.


\vspace{-1em}
\begin{acks}
We thank all reviewers for their valuable feedback. This work was funded by NSF-1845446.
\end{acks}
\bibliographystyle{ACM-Reference-Format}
\bibliography{refs.bib}

\end{document}